\documentclass[doublespacing]{elsart}

\usepackage{graphicx}
\usepackage{amssymb}
\usepackage{psfig}

\begin{document}
\bibliographystyle{unsrt}
\psfigurepath{.:figure}
\begin{frontmatter}

%----------------------------------------------------------------------
% Specify destination and version number of the manuscript

\journal{SCES'2001: Version 1}

%----------------------------------------------------------------------
% Title of manuscript

\title{Effect of La doping on magnetic structure in heavy
fermion CeRhIn$_5$}

%----------------------------------------------------------------------
% List of authors
%
% List each author using a separate \author{} command
%
% If there is more than one author address, add a label to each author
% of the form \author[label]{name}.  This label should be identical to
% the corresponding label provided with the \address command.
%
% e.g. if there are three authors from two institutions in USA and 
% France, you can link them to their respective addresses, using
%
% \author[US]{John Doe}
% \author[US,FR]{Jane Doe}
% \author[FR]{Jean Dupont}
% \address[US]{University of Life, Somewhere, USA}
% \address[FR]{Universite de la Vie, Quelque Part, France}
%
% N.B. Unlike the document class used for abstract submissions, it is
% possible to have the author associated with more than one address,
% as shown in the example above.
%

\author[LANL]{W. Bao\corauthref{1}\thanksref{ABC}}
\author[LANL]{A.D. Christianson}
\author[LANL]{P.G. Pagliuso}
\author[LANL]{J.L. Sarrao}
\author[LANL]{J.D. Thompson}
\author[LANL]{A.H. Lacerda}
\author[NIST]{J.W. Lynn}

%----------------------------------------------------------------------
% List of addresses
%
% If there is more than one address, list each using a separate 
% \address command using a label to link it to the respective author
% as described above

\address[LANL]{Los Alamos National Laboratory, Los Alamos, NM 87544, USA}
\address[NIST]{National Institute of Standards and Technology, Gaithersburg, MD 20899, USA}

%----------------------------------------------------------------------
% Title page footnotes
%
% If you need to add qualifying information to any of the authors, 
% use the \thanksref{} command within the \author command.  The 
% argument is the label of a corresponding \thanks[label]{text}
% command which contains the footnote text
%
% e.g. you can acknowledge a funding authority for John Doe, using
%
% \author{John Doe\thanksref{ABC}}
% \thanks[ABC]{This work was supported by Institute of Unphysical 
%    Phenomena under contract no. ABC-123}
%

\thanks[ABC]{Work at LANL was supported by the U.S. Dept. of Energy.}

%----------------------------------------------------------------------
% Contact Information
%
% Add the complete postal address, telephone number, fax number, and
% email address of the corresponding author as a special footnote using
% the \corauth[]{} command.  This works in a similar way to the \thanks 
% command.  Add the \corauthref{} command within the \author command.
% The argument is the label of a corresponding \corauth[label]{text}
% command which contains the contact information.  Prefix the text with
% Corresponding Author:
%
% e.g. if the contact author is John Doe,
%
% \author{John Doe\corauthref{1}}
% \corauth[1]{Corresponding Author: University of Life, 123 Some St.,
%    Somewhere, MI 12345, USA.  Phone: (555) 555-5555 
%    Fax: (555) 555-7777, Email: JDoe@uol.edu}
%

\corauth[1]{Corresponding Author: MS-K764, LANL, Los Alamos, NM 87545,
USA. Phone: (505) 665-0753, Fax: (505) 665-7652, Email: wbao@lanl.gov}

%----------------------------------------------------------------------
% Text of abstract

\begin{abstract}

The magnetic structure of Ce$_{0.9}$La$_{0.1}$RhIn$_5$ is measured using
neutron diffraction. It is identical to the incommensurate
transverse spiral for CeRhIn$_5$, with a
magnetic wave vector ${\bf q}_M=(1/2,1/2,0.297)$,
a staggered moment of 0.38(2)$\mu_B$
at 1.4K and a reduced N\'{e}el temperature of 2.7~K.

\end{abstract}

%----------------------------------------------------------------------
% Manuscript keywords
%
% Please give two or three keywords in the form: keyword \sep keyword
% e.g. NMR \sep superconductivity
%
% NB The syntax is different from the abstract document class

\begin{keyword}

magnetic structure, heavy fermion superconductor, neutron 
diffraction

\end{keyword}

%----------------------------------------------------------------------
% End of front page

\end{frontmatter}

%----------------------------------------------------------------------
% Manuscript text
%
% Fill in the following space with the manuscript text.
%
% A number of LaTeX commands may be invoked in this space, e.g.
%
% \section{} : to insert a new section title
% \label{}   : to label the numbered section for use in \ref{}
% \cite{}    : to add a reference using the label in \bibitem{}
% 
% A number of LaTeX environments may be used, e.g. 
% \begin{equation}
%     An equation inserted here will be automatically numbered
% \end{equation}  
%
% Please refer to other LaTeX documentation for help on using these
% environments.

Superconductivity has been recently discovered in heavy fermion
materials Ce$M$In$_5$ with transition temperature
$T_C=2.1$~K for $M$=Rh at 17 kbar, $T_C=0.4$~K for $M$=Ir,
and $T_C=2.3$~K for $M$=Co\cite{hegger}.
Lines of nodes in the superconducting gap have been indicated from
thermodynamic, transport and NQR measurements\cite{roman,zheng,izawa}.
This type of anisotropic superconductivity in heavy fermion materials
is widely believed to be mediated by
antiferromagnetic fluctuations. While a two-dimensional (2D) Fermi surface of
undulating cylinders is detected in the de Haas-van Alphen 
measurements\cite{haga}, 
antiferromagnetic correlations are 3D from direct measurements using 
neutron scattering\cite{bao01b} and from a theoretic fit to the 
NQR measurements\cite{zheng}.

CeRhIn$_5$ is an antiferromagnet below $T_N=3.8$~K at ambient pressure.
A spiral magnetic structure was first detected 
with the NQR measurements\cite{nqr}.
Neutron diffraction measurements reveal a magnetic moment of 0.374(5)$\mu_B$
residing on the Ce ion and forming an incommensurate spiral with a
magnetic wave vector ${\bf q}_M=(1/2,1/2,0.297)$\cite{bao00a}. This
may be contrasted with commensurate antiferromagnetic structures of 
structurally related heavy fermion Ce$_2$RhIn$_8$ and CeIn$_3$\cite{bao01a}.
Applying pressure to CeRhIn$_5$ or doping it
with Ir on the Rh site has only a small effect
on $T_N$ until the material becomes a superconductor\cite{hegger,pgpRhIr},
while doping with La on the Ce site reduces $T_N$ linearly\cite{pgpCeLa}.
We have found with neutron diffraction measurements that 
the incommensurate magnetic structure is robust against 
these external perturbations.
Results on Ce$_{0.9}$La$_{0.1}$RhIn$_5$ are reported here. 

A plate-like single crystal sample of Ce$_{0.9}$La$_{0.1}$RhIn$_5$, weighing
0.4 g, was grown from an In flux.
It has the tetragonal HoCoGa$_5$ structure (space group No. 123, P4/mmm)
with lattice parameters $a=4.643\AA$ and $b=7.530\AA$ below 20~K.
Neutron diffraction experiments were performed at NIST using the thermal 
triple-axis spectrometer BT9 in a two-axis mode. Neutrons with incident
energy $E=35$~meV were selected using the (002) reflection of a pyrolytic
graphite (PG) monochromator. PG filters of total thickness 8cm were used to
remove higher-order neutrons. 
The horizontal collimations were 40-48-48. 
The sample temperature was regulated by a top loading pumped He cryostat.

Temperature dependent Bragg peaks were found in Ce$_{0.9}$La$_{0.1}$RhIn$_5$
along the (1/2,1/2,$l$) line only at incommensurate
positions characterized by ${\bf q}_M=(1/2,1/2,0.297)$, which is
identical to that for pure CeRhIn$_5$. Fig. 1(a) shows a pair of
magnetic satellite peaks in a Brillouin zone at 1.4~K and above 
the N\'{e}el temperature. The intensity of the 
(1/2,1/2,1.297) Bragg peak is shown in Fig 1(b) as the square of the
order parameter of the antiferromagnetic phase transition. A
N\'{e}el temperature of 2.7~K is determined. 
As shown by the scans at various temperatures in the inset to Fig. 1(b),
there is no detectable temperature dependence in the incommensurate
magnetic wave vector. This is identical to what is observed 
for pure CeRhIn$_5$\cite{bao00a}.

To determine the magnetic structure for Ce$_{0.9}$La$_{0.1}$RhIn$_5$,
20 independent magnetic Bragg peaks were measured with rocking scans
at 1.4~K.
Integrated intensities of these peaks are normalized to the structural
Bragg peak (220) to yield magnetic cross sections,
$\sigma_{obs}({\bf q})=I({\bf q})\sin(\theta_4)$, 
in absolute units (see Table 1). 
The same magnetic model for CeRhIn$_5$\cite{bao00a},
\[
\sigma({\bf q})=\left(\frac{\gamma r_0}{2}\right)^2
	\langle M\rangle^2 \left|f(q)\right|^2 
	\left(1+|\widehat{\bf q}\cdot \widehat{\bf c}|^2
	\right),
\]
offers a reasonable description for Ce$_{0.9}$La$_{0.1}$RhIn$_5$. The
staggered magnetic moment, $M$, at 1.4K is 0.38(2)$\mu_B$ 
per Ce$_{0.9}$La$_{0.1}$RhIn$_5$. Within the error bar,
this is the same as the moment
for pure CeRhIn$_5$.
Calculated cross sections, $\sigma_{cal}$,
are listed in Table 1.

In summary, magnetic structure of Ce$_{0.9}$La$_{0.1}$RhIn$_5$ is identical 
to that for pure CeRhIn$_5$. While the N\'{e}el temperature is reduced
from 3.8~K to 2.7~K upon 10\% substitution of Ce by La, the staggered
moment remains the same within the error bar.

%----------------------------------------------------------------------
% Reference section
%
% List each reference with a separate \bibitem{} command.  The
% argument contains the label that is used in the \cite{} command
% in the main text
%
% e.g.
%
%    This follows our pioneering work on TdB2\cite{TdB2}.
%
% \bibitem{TdB2}
% J. Doe, J. Doe, and J. Dupont, J. Irrep. Res. 10 (2000) 1000.

%\bibliography{../../../tex/bib/heavyf,../../../tex/bib/mypaper}

%----------------------------------------------------------------------
% Figures and Tables
%
% Insert figures and tables at the end of the document  
%
% \begin{figure}
%     \centering
%     \includegraphics{filename.eps}
%     \caption{Insert figure caption here} 
% \end{figure}  
%
% \begin{table}
%     \centering
%     \begin{tabular}
%     Insert table here
%     \end{tabular}
%     \caption{Insert table caption here}
% \end{table}  
%
% Please refer to other LaTeX documentation for help on using these
% environments.

\begin{figure}
\centerline{
\psfig{file=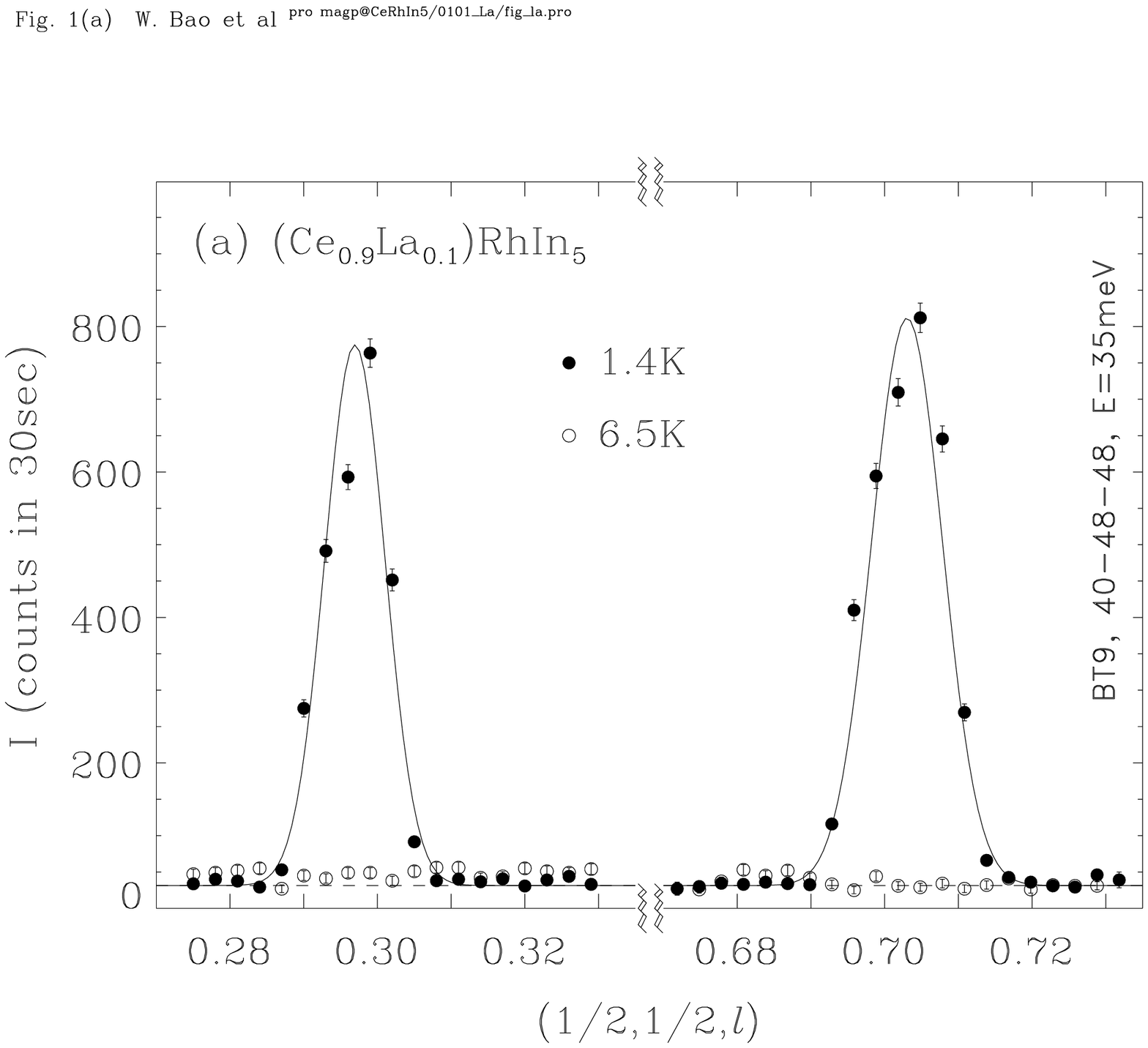,width=.8\columnwidth,angle=90,clip=}}

\centerline{
\psfig{file=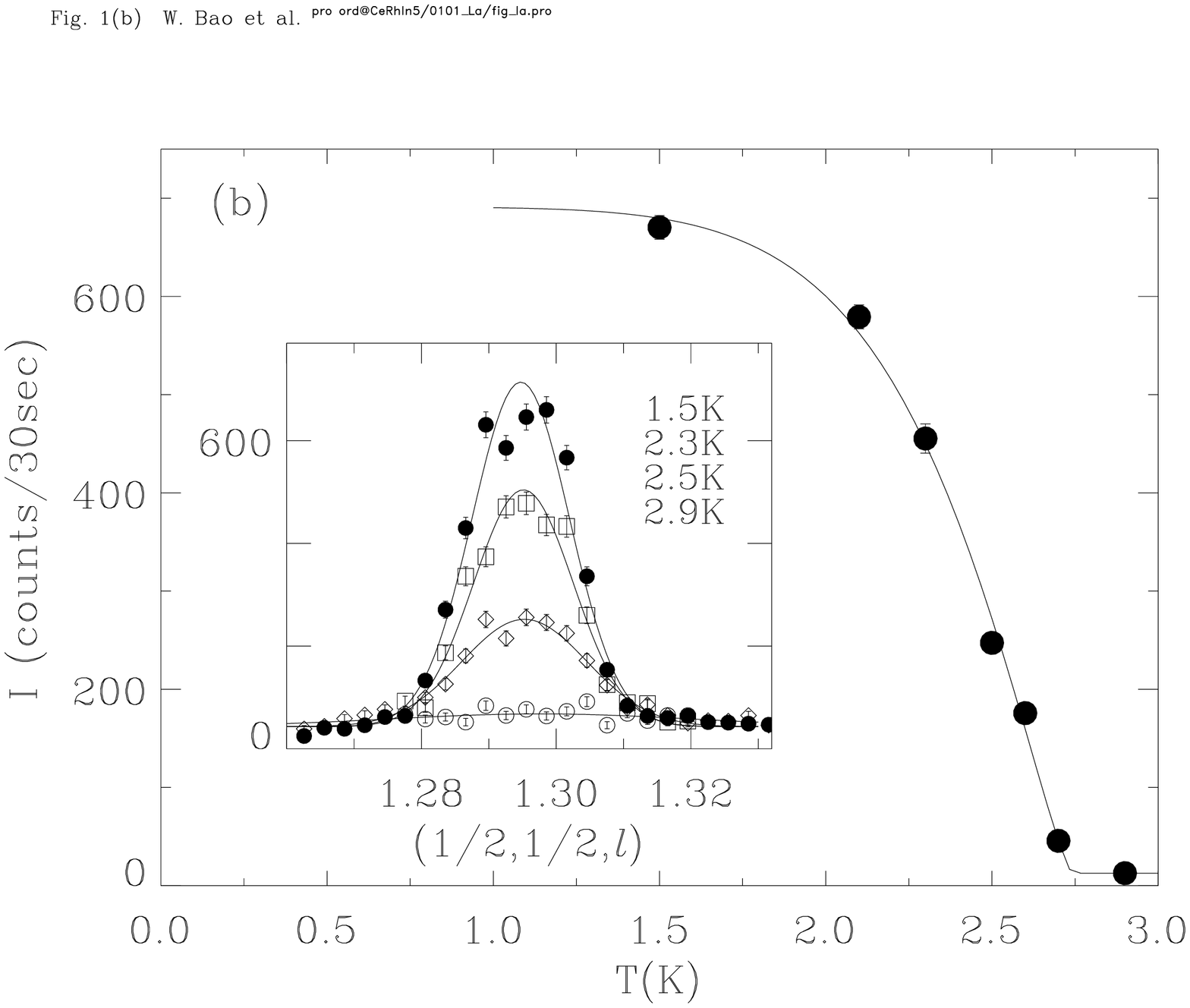,width=.8\columnwidth,angle=90,clip=}}
\caption{(a) Elastic scan through a pair of magnetic Bragg points 
at 1.4 and 6.5~K. (b) Temperature dependence of the (1/2,1/2,1.297)
Bragg peak. Inset: Elastic scans, with decreasing intensity, at
1.5, 2.3, 2.5, and 2.9~K. 
} 
\end{figure}

\begin{table}
\centering
\begin{tabular}{ccc|ccc}
\hline
\hline
${\bf q}$ & $\sigma_{obs}$ & $\sigma_{cal}$ &
${\bf q}$ & $\sigma_{obs}$ & $\sigma_{cal}$ \\
\hline
  (0.5  0.5  -0.297) &  9.9(2) & 10.0 &
  (0.5  0.5   0.297) & 12.1(2) & 10.0 \\
  (0.5  0.5   0.703) & 14.8(3) & 11.6 &
  (0.5  0.5   1.297) & 15.7(3) & 13.0 \\
  (0.5  0.5   1.703) & 14.2(3) & 12.9 &
  (0.5  0.5   2.297) & 10.0(3) & 11.6 \\
  (0.5  0.5   2.703) &  6.8(3) & 10.3 &
  (0.5  0.5   3.297) &  4.6(3) &  8.3 \\
  (0.5  0.5   3.703) &  3.7(3) &  7.0 &
  (0.5  0.5   4.297) &  3.7(2) &  5.2 \\
  (0.5  0.5   4.703) &  4.3(2) &  4.2 &
  (0.5  0.5   5.297) &  3.2(2) &  2.9 \\
  (0.5  0.5   5.703) &  2.9(2) &  2.3 &
  (0.5  0.5   6.297) &  1.8(2) &  1.5 \\
  (1.5  1.5   0.297) &  4.8(2) &  4.5 &
  (1.5  1.5   0.703) &  4.9(2) &  4.5 \\
  (1.5  1.5   1.297) &  5.2(3) &  4.5 &
  (1.5  1.5   1.703) &  5.1(2) &  4.5 \\
  (1.5  1.5   2.297) &  4.1(2) &  4.2 &
  (1.5  1.5   2.703) &  3.4(2) &  3.9 \\
  (2.5  2.5   1.297) &  0.9(1) & 1.2 &
  (2.5  2.5   1.703) &  1.2(1) &  1.1 \\
\hline
\hline
\end{tabular}
\caption{Magnetic Bragg intensity, $\sigma_{obs}$, observed at
1.4~K in unit of 10$^{-3}$b per Ce$_{0.9}$La$_{0.1}$RhIn$_5$.
The theoretic intensity, $\sigma_{cal}$, is calculated using 
$M=0.38\mu_B$ per Ce$_{0.9}$La$_{0.1}$RhIn$_5$.}
\end{table}  

%----------------------------------------------------------------------
% Terminate document

\end{document}